\documentclass[twocolumn,showpacs,prl,preprintnumbers,superscriptaddress]{revtex4}
\usepackage{graphicx}

\def\bea{\begin{eqnarray}}
\def\eea{\end{eqnarray}}
\def\a{\alpha}
\def\d{\delta}

\def\nn{\nonumber}

\def\la{\langle}
\def\ra{\rangle}

\def\o{\omega}
\def\n{\eta}
\def\g{\gamma}
\def\bFi{{\Phi}}
\def\bM{{ M}}
\def\bY{{\cal{Y}}}
\def\bG{{\Gamma}}
\def\l{\lambda}

\def\f{\frac}

\def\bd{{\bf d}}
\def\ba{{\bf a}}
\def\be{{\bf e}}
\def\bb{{\bf b}}
\def\l{\lambda}
\def\tg{\tilde{\gamma}}
\def\tx{\tilde{x}}
\def\teta{\tilde{\eta}}

\begin{document}

\title{Heat Conduction in two-dimensional harmonic crystal with disorder}
\author{Lik Wee Lee}
\affiliation{Physics Department, University of California, Santa Cruz,
CA 95064}
\author{Abhishek Dhar}
\affiliation{Raman Research Institute, Bangalore 560080}
\date{\today}
\begin{abstract}
We study the problem of heat conduction in a mass-disordered two-dimensional
harmonic crystal. Using two different stochastic heat baths, we perform
simulations to determine the system size ($L$) dependence of the heat
current ($J$). For white noise heat baths we find that $J
\sim 1/L^{\a}$  with $\a \approx 0.59$ while correlated noise heat baths
gives $\a \approx 0.51$. A special case with correlated
disorder is studied analytically and gives $\a=3/2$ which agrees also
with results from exact numerics.
\end{abstract}
\maketitle

\section{Introduction}

The problem of proving or verifying Fourier's law, 
$J = -\kappa \nabla T$ where $\kappa$ is the thermal conductivity, 
in any system evolving through Newtonian dynamics has been a challenge for theorists
\cite{bonet,lepri}. So far most studies have been restricted to
one-dimensional systems for the simple reason that they are easier to
study through simulations and through whatever analytic methods that
are available. Also the hope is that such studies on one-dimensional
systems provides insights which will be useful when one confronts the
more difficult (and experimentally more relevant) problem of higher
dimensional systems.
For one dimensional systems, some of the most interesting results that
have been obtained are:
(1) For momentum conserving nonlinear systems, the heat current $J$
decreases with system size $L$ as $J \sim 1/L^\alpha$ where $\alpha <
1$ \cite{dhar}. Thus Fourier's law (which predicts 
$\alpha = 1$) is not valid. The exponent
$\alpha$ is expected to be universal but its exact value is still
not known. A renormalization group analysis of the hydrodynamic
equations \cite{onut} predicts $\alpha=2/3$ while mode-coupling theory
\cite{lepri2} gives
$\alpha=3/5$.  The results from simulations are not able to
convincingly decide between either of these. 
(2) For disordered harmonic systems we get $J \sim 1/L^\alpha$ again 
but the exponent $\alpha$ depends on the properties of the heat baths
\cite{dhar2,connor,rubin,cash}.

In dimensions higher than one, there are few detailed studies and
it is fair to say that it is totally unclear as to whether or not
Fourier's law will hold and, if not, then what the value of the exponent
$\a$ is.  
For nonlinear systems which are expected to show local thermal
equilibrium,  both the hydrodynamic approach 
and mode coupling theories predict  a logarithmic divergence of the
conductivity in two dimensions. There have been simulations by Lippi
and Livi \cite{lippi}
who find a logarithmic divergence but simulations on larger-size
systems by Grassberger et al \cite{grass2} seem to obtain a power law
divergence. 
A disordered harmonic model in $2D$ was studied in simulations by Yang
\cite{yang} who claimed that beyond some critical disorder one gets
Fourier's law, {\emph{i.e}} $\a=1$. It is doubtful if this claim is
correct. The data in the paper seems to indicate $J \sim 1/L^2$  which
is \emph{not} Fourier's law. 
Besides, these simulations were done with Nose-Hoover
heat baths and it is known that these can be problematic when applied
to harmonic systems \cite{dhar3}. Simulations by Hu et al \cite{hu} on
the same model but with stochastic heat baths do not find a Fourier behaviour. 
Finally an older study by Poetzsch et al \cite{poet} looked at heat
conduction in a $2D$ system with both disorder
and nonlinearity and they give some evidence for Fourier behaviour. 

In this paper we consider heat conduction in a $2D$ disordered harmonic
system. Let us first try to see what one should expect theoretically. 
We expect localization phenomena (for phonons) to play an important
role. A renormalization group calculation \cite{john} predicts: in $1D$ all
modes with $\o > 1/L^{1/2}$ are localized; in $2D$  
all modes with $\o > [\log(L)]^{-1/2}$ are localized; 
in $3D$ there is a finite band of frequencies of  extended states. 
This is similar to results for electron localization with the
important difference that here the $\o \to 0$  modes are extended even in
$1D$ and $2D$. Also for the case of electrons, only electrons near the
Fermi-level contribute significantly to transport 
while in heat transport all phonons contribute. From the localization
results we expect that in 
$3D$ the current in a disordered harmonic system should be independent of
system size ($\a=0$). In one and two
dimensions it is the small number of low-frequency phonons ($\o <
\o_c$) which dominate transport properties. The fact that $\o_c \to 0$
with increasing $L$ immediately implies that $\a >0$. In $1D$ it has
been shown \cite{dhar2} that the
exact value of $\a$ depends on the low frequency spectral properties
of the bath. A similar calculation is not available in the $2D$ case
and we address this specific question.  

Here we present results from a detailed simulational study to determine the exponent
$\a$ for a mass-disordered harmonic system. Two
different kinds of stochastic baths are considered, one with white noise and
the other with correlated noise. We also study a special case where the
disorder is correlated. 

{\bf Definition of model}:
We consider heat conduction in a two-dimensional mass-disordered
harmonic crystal described by the Hamiltonian 
\bea
&H&= \sum_{\stackrel{i=1,L_x}{j=1,L_y}} \f{p_{ij}^2}{2 m_{ij}} +
\sum_{\stackrel{i=1,L_x}{j=1,L_y}} \f{1}{2}
    [(x_{ij}-x_{i-1j})^2 \nn \\&+&(x_{ij}-x_{i+1j})^2   +
(x_{ij}-x_{ij-1})^2+(x_{ij}-x_{ij+1})^2] \nn
\eea
where $\{x_{ij},p_{ij},m_{ij} \}$ denote the position (particle
displacements about equilibrium positions), momentum and mass of a
particle at the site $(i,j)$. 
We set the masses of exactly half the particles to
one and the remaining to two and make all  configurations equally probable. 
Heat conduction takes place
in the $x$-direction and we assume that the ends of the system are
fixed by the boundary conditions $x_{0j}=0=x_{L_x+1j }$. 
We will assume periodic boundary conditions along the
$y$-direction so that $x_{i j+L_y}=x_{ij}$. 
The heat baths are modeled through Langevin equations and thus we get the
following equations of motion:
\bea
&&m_{1j} \ddot{x}_{1j}=-4 x_{1j}+x_{2j}+x_{1j-1}+x_{1j+1} + h^L_j \nn \\
&&m_{L_xj} \ddot{x}_{L_xj}=-4
x_{L_xj}+x_{L_x-1j}+x_{L_xj-1}+x_{L_xj+1}+h^R_j \nn \\ 
&&m_{ij} \ddot{x}_{ij}=-4
x_{ij}+x_{i-1j}+x_{i+1j}+x_{ij-1}+x_{ij+1} 
\label{eqnmot} 
\eea
(for $1<i<L_x$ and $1 \leq j \leq L_y$),~where
$h^L_{j}$ and $h^R_j$ denote the forces from the heat baths. We will
consider two different models for the heat baths:

{\bf (I)} {\emph{Gaussian white noise source}}. Thus
$h^L_{j}=-\gamma \dot{x}_{1j}+\eta^L_{j};~  
h^R_j=-\gamma \dot{x}_{L_xj}+\eta^R_{j} $,
where the noise terms have the properties $\la \eta^L_{j} \ra=\la
\eta^R_{j} \ra=0$,~ 
$ \la \eta^L_{j}(t) \eta^L_{j'}(t') \ra =2 T_L \gamma \d (t -t') \d_{j j'},~ 
\la \eta^R_{j}(t) \eta^R_{j'}(t') \ra =2 T_R \gamma \d (t -t')
\d_{j j'}$.

{\bf{(II)}} {\emph{Gaussian exponentially correlated source}}. In this case the
bath forces have the forms:
\bea
h^L_{j}&=&-\int_{-\infty}^t dt' \tg (t-t') \dot{x}_{1j}(t')+\eta^L_{j}  \nn \\
h^R_j&=&-\int_{-\infty}^t dt' \tg (t-t') \dot{x}_{L_xj}(t')+\eta^R_{j} 
\eea
with $ \la \eta^L_{j}(t) \eta^L_{j'}(t') \ra = T_L \tg (t -t') \d_{j
  j'}$, $\la \eta^R_{j}(t) \eta^R_{j'}(t') \ra =T_R \tg (t -t')
\d_{j j'}$ and $\tg(t)= e^{-\g t}$.
A simple way of implementing correlated baths in the simulations is by
introducing  new dynamical variables $y^L_j,~y^R_j$ for the
bath and setting $h^L_j=y^L_j,~h^R_j=y^R_j$. These satisfy the equations of motion
$\dot{y}^L_j =-\g y^L_j-\dot{x}_{1j}+\eta^L_j$, etc.
In the long time limit it can be easily seen that the solutions
$y^L_j(t)$  and $y^R_j(t)$ have the required properties of correlated baths.

We now discuss the results from simulations of the two different bath
models.

\begin{figure}
\includegraphics[width=3in]{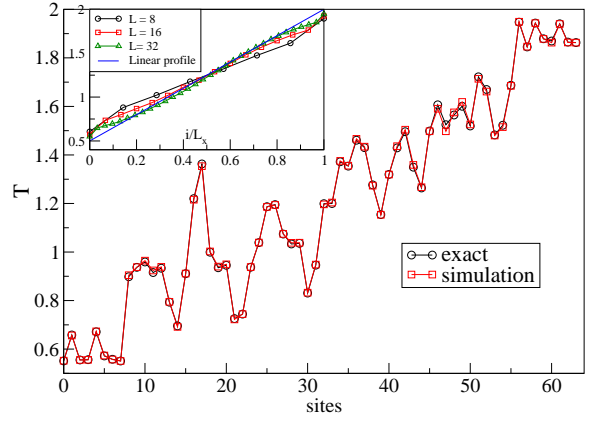}
\caption{Temperature at all the sites of a $8\times 8$ fully disordered
  lattice, from simulations and from the exact solution. Inset shows
  the disorder-averaged temperature profiles for different system sizes and seems to
  approach a linear form.}
\label{exact} 
\end{figure}

{\bf Simulations with white noise}: Equilibration times in simulations of
disordered harmonic lattices can be very long and this can sometimes lead to
wrong conclusions [see for e.g \cite{dharcom}]. To avoid such problems
we first compare our simulation results with exact numerical results
on steady state properties of small systems. We now briefly describe the
numerical technique.   

With $N=L_xL_y$ let us define the new 
variables $\{ q_1,q_2,...,q_{2N} \}= \{
x_{11},x_{12}...,x_{L_xL_y},p_{11},p_{12}...,p_{L_xL_y} \}$. Then
Eq.~\ref{eqnmot} can be rewritten in the form: 
\bea
\dot{q}_l=-\sum_{m=1}^{2N} \ba_{lm}q_m+\xi_l 
\label{neqmot}
\eea
where the vector $\xi$ has all elements zero except $\xi_{N+j}=\n^L_{j};~
\xi_{2 N-L_y+j}=\n^R_{j}$ (for $1 \leq j \leq L_y$)  and the $2N \times 2N$ matrix $\ba$ is given by:
\bea
 \ba= \left( \begin{array}{cc}
0 & -\bM^{-1} \\ \bFi & \bG \end{array} \right) \nn 
\eea
where the $N\times N$ matrices $\bM,\bFi,~\bG$ can be labeled by the
double indices $(i,j)$ and are given by
\bea
&& \bM_{(ij),(i'j')}= \delta_{ii'} \d_{jj'} m_{ij};~  
\bFi_{(ij)(i'j')}=4\delta_{ii'} \delta_{jj'} \nn \\
&&~~~~~~ -\d_{i
  i'}(\d_{jj'-1}+\d_{jj'+1})-\d_{j j'} (\d_{i i'-1}+\d_{i i'+1}) \nn \\
&& \bG_{(ij)(i'j')}=\g \delta_{ii'}\d_{jj'} (
\d_{i1}/m_{1j}+\delta_{iL_x} /m_{L_xj} ). 
\label{defn}
\eea
In the steady state $\la d(q_nq_l)/dt \ra =0$. From this and using 
Eq.~\ref{neqmot} we get the matrix equation \cite{ried}
\bea
\ba \bb+\bb \ba^T= \bd= \left( \begin{array}{cc}
0 & 0 \\ 0 & \be \end{array} \right)   
\label{mateq}
\eea
where $\bb$ is the correlation matrix with elements $\bb_{nl}=\la q_n q_l
\ra$ and $\be_{(ij)(i'j')}=\g \d_{ii'} \d_{jj'} (2 T_L \delta_{i1}+2 T_R \delta_{i
L_x })$. Inverting this equation one obtains $\bb$ and thus all the
moments which includes the local temperatures $T_{ij}=\la p_{ij}^2/m_{ij} \ra$ and
currents $J^x_{ij}=\la x_{i-1j} p_{ij}/m_{ij} \ra$. The dimension
of the matrix which has to be inverted is $N(2N+1) \times N(2N+1)$ and
using the fact that it is a sparse matrix we have been able to
numerically \cite{superlu} obtain $\bb$ for system sizes up to $L_x=L_y=L=8$.

The molecular dynamics simulations were performed using a
velocity-verlet scheme \cite{allen}. We chose a step-size of $\Delta
t=0.005$ and averaged over $10^8$ time steps
(for $L=256$ we took $\Delta t=0.02$ and $10^7-5\times 10^7$ time
steps). The temperatures at the two ends were set to $T_L=0.5$ and
$T_R=2.0$. In Fig.~\ref{exact}, we plot $T_{ij}$ at
every site, as obtained from the simulations and from
the exact solution, for a particular realization of disorder.  
The agreement is clearly very good. 
We also find that that current fluctuations decay faster than
fluctations of the local temperature. This is because, in the harmonic
model, \emph{decay of fluctuations takes place only through
coupling to the reservoirs} and
this is weak for localized modes that contribute to the
temperature. This means that equilibrated values for the current can
be obtained in smaller simulation runs.  

\begin{figure}
\includegraphics[width=3in]{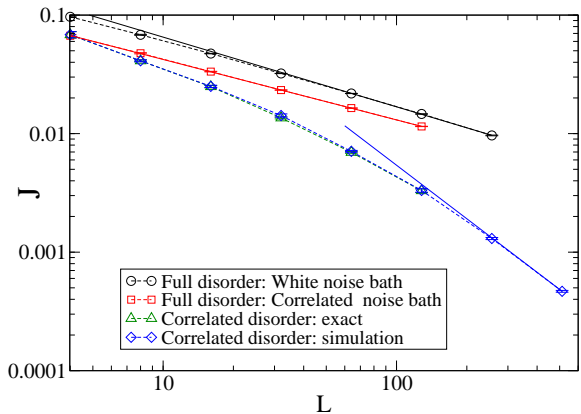}
\caption{Plot of disorder-averaged-current versus system size for the
  two different heat baths and for the case of correlated disorder
  with white noise. For the full disorder cases, the solid lines are
  fits to the last three points and have slopes $0.59$ and $0.51$.
  For the case of correlated disorder, the slope from exact numerics
  (and also simulations) is compared to $1.5$ which is what one expects analytically.}
\label{jvsl} 
\end{figure}

Simulations were performed for sizes $L_x=L_y=L=4,8,16,...,256$ and in
Fig.~\ref{jvsl} we plot the system size-dependence of the current,
averaged over $100$ samples ($9$ samples for $L=256$). Error-bars shown
are those calculated from the disorder averaging, the thermal ones being
much smaller. 
For larger system sizes we find that we need to average over a smaller
number of realizations since the {\emph{rms}} spread in the current decreases
rapidly. From our data we estimate  $\a = 0.59 \pm 0.01$.

We briefly note that Eq.~\ref{mateq} can be solved exactly for the
ordered case, using methods similar to those in \cite{ried}. 
The current is independent of system size and given by
\bea
J&=& \f{(T_L-T_R)}{4 \pi \g}
\int_0^{2 \pi} dq \phi_1(q) \nn  
\eea
where  $\phi_1(q)=1+\f{1}{2} (\g^{-2}+\l_q)-\f{1}{2}[4  (\g^{-2}+\l_q)+
  (\g^{-2}+\l_q)^2]^{1/2} $ and $\l=2[1-\cos(q)]$. The temperature in
the bulk of the system takes the constant value  $T=(T_L+T_R)/2$. 
 
{\bf Simulations with correlated noise}:
In this case the simulations were done
by using a slightly modified version of the velocity-verlet algorithm
with a step-size $\Delta t=0.001$ and averaging over $10^8$ time steps. The
accuracy of the algorithm was tested in one-dimensions where exact
numerical results are available \cite{dhar2}. Simulations were
performed for sizes $L=4,8,16,...,128$ with disorder averages over $100$
samples ($22$ for $L=128$). The results are plotted in Fig.~\ref{jvsl}
and we estimate the exponent $\alpha = 0.51 \pm 0.01$ in this case,
which is somewhat different from the slope of $\alpha \approx 0.59$ for
the case of uncorrelated noise. It is  possible that the small
difference is a finite size
effect and for larger system sizes we  might see the same exponent.  
The error bars given are statistical
errors while those from finite-size effects are more difficult to
estimate. The next example throws some light on this aspect.

{ \bf{ Correlated disorder}}: Finally we consider a special case of
  correlated disorder (with white noise baths) which was discussed in \cite{connor}. This case
  is analytically tractable and gives us some  insights on
  possible finite size effects. 
In this model, in a given column, say the $i$th, all particles have the same
mass $m_i$. 
This case can be reduced to an effective one-dimensional problem \cite{connor}.  
Using the fact that there is order in the transverse direction,
we transform to new variables using  an orthogonal basis $\psi_j(q)$ which
satisfies the equation 
$2 \psi_j(q)-\psi_{j-1}(q)-\psi_{j+1}(q)= \lambda_q \psi_j(q)$.
We choose the $\psi_j(q)$ to be real and find that
$\lambda_q=2(1-cos(q))$ with $q=2 s \pi/L_y$ where $s=1,2,...,L_y$. The new
variables $ x_i(q)=\sum_j x_{i,j} \psi_j(q) $
satisfy the following equations of motion:
\bea
m_1 \ddot{x}_{1}(q)&=&-\mu(q) x_{1}+x_{2}-\gamma
\dot{x}_{1}+\eta^L(q) \nn \\
m_{L_x} \ddot{x}_{L_x}(q)&=&-\mu(q) x_{L_x}+x_{L_x-1}-\gamma
\dot{x}_{L_x}+\eta^R(q) \nn \\
m_i \ddot{x}_{i}(q)&=&-\mu(q)
x_{i}+x_{i-1}+x_{i+1} 
\label{eqnmot2} 
\eea
(for $1<i<L_x$), where $\mu(q)=2+\l_q$.
The transformed noise variables $\eta^L(q,t)=\sum_j \eta^L_{j}(t)
\psi_j(q)$ satisfy $\la \eta^L (q,t) \eta^L (q',t') \ra= 2 T_L \g
\d(t-t') \d_{qq'}$ and similarly for $\eta^R(q,t)$. Thus for
every $q$ we have an equation 
identical to  that of a one-dimensional disordered chain with an
additional on-site potential $V=\l_q x^2/2$.
The heat current in terms of the transformed variables is:
\bea
J =
 \f{1}{L_y} \sum_q \la (-\g \dot{x}_1(q) +\eta^L(q) ) \dot{x}_1(q)
\ra 
\label{Jcor} 
\eea  
Fourier transforming Eq.~\ref{eqnmot2} using: $ \tx_i(q,\o) =\int dt x_i (q,t) e^{-i
  \o t},~\teta^{L,R}(q,\o)=\int dt \eta^{L,R}(q,t) e^{-i \o t}$ we  get the
set of equations:
 \bea
[\mu(q)-m_1 \o^2+i \g \o]\tx_1-  \tx_2 &=&\teta^L \nn \\
- \tx_{i-1}+[\mu(q)-m_i \o^2 ]\tx_i- \tx_{i+1} &=& 0 \nn \\
- \tx_{L_x-1}+[\mu(q)-m_{L_x} \o^2+i \g \o]\tx_{L_x} &=& \teta^R \nn 
\eea
which in matrix notation can be written as 
$\bY(q,\o) \tx(q,\o)=\teta(q,\o) $
where $\tx$, $\teta$ are the vectors $( \tx_1(q,\o),...,\tx_{L_x}(q,\o))^T$
and $(\teta^L(q,\o),0,0,...,\teta^R(q,\o))^T$. The matrix $\bY=k\bFi-\o^2 \bM+i \o
\bG$ where $\bM_{nl}= \delta_{nl} m_n;~\bFi_{nl}=\mu(q) \delta_{nl}-\delta_{n
l-1}-\delta_{nl+1}$ and $\bG_{nl}=\g \delta_{nl}(
\delta_{n1}/m_1+\delta_{nL_x} /m_{L_x} )$.
After some manipulations the current in Eq.~\ref{Jcor} simplifies to give:
\bea
J=\f{\g^2(T_L-T_R)}{\pi L_y} \sum_q \int_{-\infty}^{\infty} d \o \o^2
|\bY^{-1}_{1L_x}(\o,q)|^2  
\label{meth2j}
\eea
The inverse element is given by \cite{dhar2}:
$ | \bY^{-1}_{1L_x}(\o,q)|^2  = | Det[\bY]|^{-2}$ where
$Det[\bY]= D_{1,L_x}+i \g \o (D_{2,L_x}+D_{1,L_x-1}) -\g^2 \o^2 
D_{2,L_x-1}$  
and $D_{i,i'}$ is defined to be  the determinant of
the submatrix of   $k\hat{\bFi}-\o^2 \bM$ beginning with the $i$th
row and column and ending with the $i'$th row and column. These matrix
elements can be expressed in terms of products of random matrices
\cite{dhar2}. 
Using these results one can very efficiently compute the integral in
Eq.~\ref{meth2j} and obtain $J$ accurately for quite large system
sizes ($L=512$). In Fig.~\ref{jvsl} we show the system size dependence of 
the current as obtained from the exact numerical method and also from simulations. They
agree very well and give $\a \approx {1.5}$. This value can be understood
analytically by noting that the leading contribution to the current in
Eq.~\ref{meth2j} comes from the $q\to 0$ term (finite $q$ modes decay
exponentially with system size) and this is identical to
a pure $1D$ chain for which $\a=3/2$.  

The fact that the simulation results agree extremely well with the
exact numerical results (for sizes up to $L=128$) proves the accuracy of our
simulations. Further we see that for the correlated disorder case the
asymptotic result for the exponent can already be seen  
at around $L=512$. This gives us confidence that for
the case of the fully disordered lattice we might already be close to the
asymptotic value. This is also supported by the fact that the change in slope
of the $J$-vesus-$L$ curves in Fig.~\ref{jvsl} over the system sizes studied is very small.

{\bf Conclusions:}
We have performed extensive simulations of heat conduction in a mass-disordered
harmonic solid in $2D$ which give exponents $\a \approx 0.59$ for white noise
heat baths and
$\a\approx 0.51$ for correlated noise baths.  A system with correlated
disorder gives, somewhat surprisingly,  a larger exponent $\a=3/2$. 
The combination of simulations and exact 
numerics  gives us confidence on the accuracy of our
results and also  additional insight. 
Some interesting open problems are the exact determination of the
exponent $\a$ in $2D$, for any heat bath model,  and an analytical
understanding of dependence of $\a$ on bath properties. 

We acknowledge A. P. Young for a critical reading of the manuscript.
LWL acknowledges O. Narayan and T. Mai for useful discussions.
The work of one of us (LWL) is supported by the NSF
under grant DMR 0337049. We are grateful to the Hierarchical
Systems Research Foundation (HSRF) for the usage of their computing
cluster.

\end{document}